\documentstyle[12pt,aasms4]{article}  
\def\etal{{\it et~al.~}}
\def\bsax{{\it BeppoSAX~}}

\def\asca{{\it ASCA~}}

\def\erg{~{\rm erg~ cm}^{-2}\ {\rm s}^{-1}~}

\begin{document}

\newcommand{\lessim}{\ \raise -2.truept\hbox{\rlap{\hbox{$\sim$}}\raise5.truept
	\hbox{$<$}\ }}			

\title{Hard X-ray emission from the galaxy cluster A3667}

\author{Roberto Fusco-Femiano} 
\affil{Istituto di Astrofisica Spaziale, C.N.R., via del Fosso del Cavaliere, 
I-00133 Roma, Italy - dario@saturn.ias.rm.cnr.it}
\author{Daniele Dal Fiume$^{\dagger}$}
\affil{TESRE, C.N.R., via Gobetti 101, I-40129 Bologna, Italy}
\author{Mauro Orlandini}
\affil{TESRE, C.N.R., via Gobetti 101, I-40129 Bologna, Italy}
\author{Gianfranco Brunetti}
\affil{Dipartimento di Astronomia, Univ. Bologna, via Ranzani 1, 
I-40127 Bologna, Italy}
\author{Luigina Feretti}
\affil{Istituto di Radioastronomia, C.N.R., via Gobetti 101, I-40129 Bologna, 
Italy}
\author{Gabriele Giovannini}
\affil{Istituto di Radioastronomia, C.N.R., via Gobetti 101, I-40129 Bologna, 
Italy}
\affil{Dip. di Fisica, Univ. Bologna, via B. Pichat 6/2, I-40127 Bologna, 
Italy}

$^{\dagger}$ We honour our late colleague for his invaluable contribution
to the search for nonthermal emission in clusters of galaxies. 

\begin{abstract}

We report the results of a long \bsax observation of Abell 3667, one 
of the most 
spectacular galaxy clusters in the southern sky. A clear detection of 
hard X-ray radiation up to $\sim$35 keV is reported, while a hard 
excess above the thermal gas emission is present at a marginal level that
should be considered as an upper limit to the presence of nonthermal X-ray
radiation. 
The strong  
hard excesses reported by \bsax in Coma and A2256 and the only marginal 
detection
of nonthermal emission in A3667 can be 
explained in the framework of the inverse Compton model.  
We argue that the 
nonthermal X-ray detections in the PDS energy range are related to the radio
index structure of halos and relics present in the observed clusters of
galaxies. 
\end{abstract}


\section{ Introduction}
Abell 3667 is one of the most intriguing clusters of galaxies. It contains 
one of the largest radio sources in the 
southern sky with a total extent of $\sim 30^{\prime}$ which corresponds
to $\sim 2.6h_{50}^{-1}$ Mpc. This diffuse radio emission is located to the 
north-west, well
outside the central core of the cluster. A similar but weaker radio region 
is present 
also to the south-east, yielding a very peculiar structure not observed in any
other cluster (Robertson 1991; Rottgering \etal 1997). The Mpc-scale 
radio relics may be  
originated by the ongoing merger visible in the optical (Sodre' \etal 1992), in
X-ray (Knopp, Henry \& Briel 1996) and in the weak lensing mass map 
(Joffre \etal 2000). In the optical, A3667 shows strong galaxy concentrations
around the two brightest D galaxies (Proust \etal 1988;
Sodre' \etal 1992). The main optical component is coincident with the peak 
of the X-ray emission. The \asca observation reports an average gas temperature
of 7.0$\pm$0.6 keV with a constant radially averaged profile up to
$\sim 22'$. The \asca temperature map shows that the hottest region is in
between the two groups of galaxies confirming the merger scenario 
(Markevitch, Sarazin \& Vikhlinin 1999).
Recently, a Chandra observation
reported a sharp X-ray brightness
edge near the cluster core (Vikhlinin, Markevitch \& Murray 2000) due to
a large cool gas cloud moving through the hot intracluster medium (ICM).

Major cluster mergers can produce large-scale
shocks and turbulence in the ICM establishing conditions for 
in-situ particle reacceleration and magnetic field amplification 
(Tribble 1993; Longair 1994, and references therein). The nonthermal
electron population produces emission in various energy bands. Extended 
radio sources are the most evident signatures of nonthermal processes 
in clusters
of galaxies. More recently, the detections in the Coma cluster 
(Fusco-Femiano \etal 1999; Rephaeli, Blanco \& Gruber 1999) and A2256
(Fusco-Femiano \etal 2000)
of a hard X-ray excess with respect to the thermal emission represent a 
further evidence of nonthermal
processes in some clusters of galaxies. This allows to derive additional
information on the physical conditions of the ICM environment, which cannot
be obtained by studying the thermal plasma emission only.  

Both the clusters where nonthermal X-ray emission has been 
detected show extended
radio halos in the center. Moreover a relic region is present in A2256,  
 at $\sim 8'$
from the center, more extended and brighter than the halo source 
(Bridle \etal 1979). A marginal evidence
of nonthermal radiation is reported in the external regions of
the MECS detector for A2199 (Kaastra \etal 1999),  
a cluster which does not show any extended 
radio region. 
So, the most direct explanation for the detected hard X-ray excesses is
that they are due to inverse Compton (IC) scattering of 
cosmic microwave background
(CMB) photons by the relativistic electrons responsible for the radio emission.
Alternative interpretations to the IC model for the nonthermal 
radiation detected
in the Coma cluster have been proposed. Blasi \& Colafrancesco (1999) 
suggest a secondary
electron production. However, this model implies a $\gamma$-ray flux 
considerably larger
than the EGRET upper limit unless the hard X-ray excess and the radio halo
emission in Coma are not due to the same population of electrons. A different 
mechanism is given by nonthermal
bremsstrahlung from suprathermal electrons formed through the current
acceleration of the thermal gas (Ensslin \etal 1999; Dogiel 2000; 
Sarazin \& Kempner 2000;  
Blasi 2000). At present, due to the low efficiency of the proposed
acceleration processes and 
of the bremsstrahlung mechanism, these models would require an unrealistically 
high energy input as recently pointed out by 
Petrosian (2001). Besides,  
this interpretation seems to be in conflict with the \bsax detection 
of A2256, since 
the 
derived flat power-law spectrum of the electrons gives  
a negligible nonthermal bremsstrahlung contribution to the PDS 
flux (Fusco-Femiano \etal 2000).
These alternative models are motivated by the discrepancy
between the value for the intracluster magnetic field derived by the 
\bsax observation of the Coma cluster ($B_{XR}\sim 0.16\mu G$, Fusco-Femiano 
\etal 1999)
and the value derived from Faraday rotation (FR) of polarized radiation
toward the radio galaxy NGC4869 ($B_{FR}\sim 6\mu G$, Feretti \etal 1995). 
However,
this discrepancy could be resolved considering that Feretti \etal (1995) 
inferred also the existence of a weaker and larger
scale magnetic field in the range 0.1-0.2 $h_{50}^{1/2}~\mu G$ consistent
with the value measured by \bsax, and therefore the component of $\sim 6\mu G$
could be local. 
Goldshmidt \& Rephaeli (1993) suggested that the discrepancy
between $B_{XR}$ and $B_{FR}$ could be alleviated taking into consideration
the expected spatial profiles of the magnetic field and relativistic electrons.
More recently, it has been shown that IC models which include the effects of 
more realistic electron
spectra, combined with the expected spatial profiles of the magnetic field, and 
anysotropies in the pitch angle distribution of the electrons 
allow higher values of the intracluster magnetic field in better agreement
with the Faraday rotation measurements (Brunetti \etal 2001; Petrosian 2001).  
  
In this Letter, we present the results of a long observation of A3667,
exploiting the unique capabilities of the PDS (Frontera \etal 1997) onboard 
\bsax, to search for
hard X-ray radiation (HXR) emission. Besides, we discuss a possible explanation for the 
marginal detection of the nonthermal hard excess with respect to the thermal
emission reported by this observation. 

Throughout the Letter we assume a Hubble constant of 
$H_o = 50~km~s^{-1}~Mpc^{-1}~h_{50}$ and
$q_0 = 1/2$, so that an angular 
distance of $1^{\prime}$ corresponds to 87 kpc ($z_{A3667} = 0.055$, Sodre'
\etal 1992). 
Quoted confidence intervals are at $90\%$ level, if not otherwise specified.

\section {PDS Data Reduction and Results}

The pointing coordinates of \bsax are at J(2000): $\alpha:~20^h~ 11^m~ 30.^s$; 
$\delta:~ -56^{\circ}~ 40'~ 0''.0$,
in proximity of the secondary galaxy concentration and the PDS FOV
(1.3$^{\circ}$) includes only the radio region in the north of the cluster.
The total effective exposure time for the PDS was
$\sim 1.13\times 10^5$ sec  
in the two observations of May 1998 and October 1999.

Since the source is rather faint in the PDS band (approximately 0.44 mCrab 
in 15-40 keV) a careful check of the background subtraction must be performed. The background sampling
was performed using the default rocking law of the two PDS collimators that 
samples ON, +OFF, ON, -OFF fields for each collimator with a dwell time of 
96 sec (Frontera et al. 1997). When
one collimator is pointing ON source, the other collimator is pointing 
toward one of the two OFF positions. We used the standard procedure to obtain 
PDS spectra (Dal Fiume et al. 1997); this procedure consists of 
extracting one accumulated spectrum for each unit for each collimator 
position. 
We then checked the two independently accumulated background spectra in the 
two different +/-OFF sky directions, offset by $210'$ with respect to the 
on-axis pointing direction. 
The comparison between the two accumulated backgrounds 
([+OFF] vs. [-OFF]) does not show difference in the two pointings. 
The background level of the PDS is the lowest and more stable obtained so far 
with the
high-energy instruments on board satellites thanks to its equatorial orbit.
No modeling of the time
variation of the background is required. The correctness of the PDS
background subtraction has been checked by verifying that the counts fluctuate
at about zero flux as the signal falls below detectability. This happens at
energies greater than $\sim$35 keV.

Figure 1 shows a clear detection of hard X-ray emission up to $\sim$35 keV,
at a confidence level of $\sim 10\sigma$. The observed count rate was 
0.134$\pm$0.013 cts/s. The fit to the PDS data with a thermal component  
(continuous line) at the fixed average gas temperature of 7 keV
derived from the \asca observation (see Introduction) indicates a marginal  
presence of a hard excess at a confidence level of $\sim 2.6\sigma$.
If we introduce a second nonthermal component, modeled as a power law, 
we obtain an insignificant improvement with respect to the previous model, 
according to the F-test. 
The low statistics do not allow us to perform a fit with a thermal component
at the fixed average cluster gas temperature and a 
second thermal component. The only possibile  
fit is with a thermal bremsstrahlung model that gives a best
fit temperature of $\sim$ 11 keV with a large confidence interval 
(7.7-16.8 keV, 90\%). The average cluster 
gas temperature is only slightly out of this interval. 
Therefore, the $\sim 2.6\sigma$ excess should be considered as an upper limit
to the presence of nonthermal HXR radiation. 
The analysis of the two
observations with effective exposure times of $\sim$44 ks (1998 May) and
$\sim$69 ks (1999 October) for the PDS does not show significant 
flux variations.

\section{Discussion}

The long observation by \bsax of Abell 3667 reports a robust evidence for
hard X-ray emission in the PDS energy range 15-35 keV and an upper limit 
for a hard nonthermal excess with respect to the thermal emission 
at the average gas
temperature of 7 keV.

As pointed out in the Introduction, the most striking feature of Abell 3667
is the diffuse arc-shaped radio region located to the NW of the X-ray core
with a total extent of  $\sim 30^{\prime}$
($\sim$ 2.6 Mpc), one of the largest radio sources known in the southern sky.
The total flux density is 5.5$\pm$0.5 Jy at 843 MHz. Also the spectral
index structure is very interesting: a region with a flat spectrum
($\alpha_r\sim$ 0.5) in the NW rim of the source with a considerable steepening
to $\alpha_r\sim$1.5 toward the SW (Rottgering \etal 1997).
The overall radio
spectral index of $\sim$1.1
is consistent with the flux density measurements (Mills \etal 1961; Bolton,
Gardner \& Mackey 1964; Robertson 1991).
Fitting the PDS data with a thermal component, at the temperature of 7 keV, 
and a power law component, with index 2.1 (1+$\alpha_r$),
we derive a nonthermal IC flux upper limit of 
$\sim 4.1\times 10^{-12}\erg$
in the 15-35 keV energy. Extrapolating this flux in the energy range
20-80 keV we obtain $F_X\sim 6.4\times 10^{-12}\erg$
that is a factor $\sim$3.4 and $\sim$2 lower than the nonthermal fluxes
detected
in Coma and A2256, respectively. In the IC interpretation this flux
upper limit,
combined with the radio synchrotron emission, determines a lower
limit to the volume-averaged intracluster magnetic field of 0.41$\mu G$.

Given the presence of such a large radio region in the NW of A3667, a robust 
detection of a nonthermal X-ray component might be expected instead of the
upper limit reported by \bsax. This result cannot be attributed to
the shift
($\sim 17^{\prime}$)
 of the PDS
pointing with respect to the centroid of the radio relic
($\alpha: 20^h~ 10^m~ 30^s.0$; $\delta: -56^{\circ}~ 25'~ 0.''0$). 
Instead, 
one possibile explanation 
may be related to 
the radio spectral index structure of the NW relic. As indicated
by Roettiger, Burns \& Stone (1999) the  
sharp edge of the radio source is the site of particle acceleration, while
the progressive index steepening with the increase of the distance from
the shock would indicate particle aging
because of radiative losses. In the narrow shocked region, where particle
reacceleration is at work,
the magnetic field is expected
to be amplified by adiabatic
compression, with the consequence that the synchrotron emission is enhanced, 
thus
giving a limited number of electrons the ability to produce IC X-rays.
In the postshock region of the relic,
the electrons suffer strong radiative losses with no reacceleration, 
considering also that the relic
is well outside the cluster core. Therefore, their energy spectrum develops
a high energy cutoff at $\gamma < 10^4$, and the electron energy is not
sufficiently high to emit IC radiation in the hard X-ray band. Synchrotron
emission is detected from the postshocked region, because the magnetic field
is still strong due to its likely long relax time. 

The Coma cluster, where nonthermal HXR 
emission is present at a significant level (Fusco-Femiano \etal 1999),
shows a quite different radio index structure
(Giovannini \etal 1993; Deiss \etal 1997). The cluster exhibits a central 
plateau (R$\sim 10'$) with radio spectral index $\sim$0.7 (in the core, it
appears to be lower; Deiss \etal 1997) and a
progressive spectral steepening
with the increasing radius in the external regions of the radio halo. Moreover, 
the total extent of the radio
halo (R$\sim 80'$) and the lack of a clear shocked region are not 
compatible with the scenario of the pure spectral aging
of the emitting electrons.
Thus, in situ reacceleration processes are required, probably because of 
turbulence related to recent mergers 
(Colless \& Dunn 1996, Donnelly \etal 1999; Arnaud \etal 2001)
with a possible additional contribution from the
 gas motion originating from
the massive galaxies orbiting in the cluster core
(Deiss \& Just 1996).
Brunetti \etal (2001) have recently shown that
the radio observational properties of the Coma halo and the nonthermal HXR
emission 
can be accounted for by a population of reaccelerated
relativistic electrons with energy break $\gamma_b $
\raise 2pt \hbox {$>$} \kern-1.1em \lower 4pt \hbox {$\sim$}
$10^4$ emitting in a magnetic field
smoothly decreasing from the center toward the
periphery. 

Similar considerations can be applied to Abell 2256, that
is the second cluster that shows a clear evidence of a hard excess above
the thermal intracluster emission
(Fusco-Femiano \etal 2000).
Abell 2256 exhibits a
large diffuse radio region (1.0$\times$0.3 Mpc) in
the north, at a
distance of $\sim 8^{\prime}$ from the cluster center (relic),
with a rather uniform and flat spectral index of 0.8$\pm$0.1
between 610 and 1415 MHz. A fainter extended emission (halo)
permeates the cluster
center with a steeper radio spectral index of $\sim$1.8
(Bridle \& Fomalont 1976; Bridle \etal 1979;
Rottgering \etal 1994; Rengelink \etal 1997).
The low and uniform value of $\alpha_r$ in the cluster
relic indicates a broad reacceleration region, probably the result of an 
ongoing merger event (Sun \etal 2001). 
Moreover,
the presence of the central radio halo would favour
the hypothesis that in situ reacceleration processes
are active in the cluster volume.
As discussed by Fusco-Femiano  \etal (2000),
in the framework of the IC model, the probable source of the
non-thermal HXR emission is the large relic if the
associated magnetic field is of the order of $\sim 0.1$ $\mu$G.
An additional contribution could be provided by the halo
electrons if the radio index structure is similar to that of the Coma cluster.

\section{Conclusions}

The positive detections of nonthermal HXR radiation in the Coma cluster
and A2256 and the upper limit reported in A3667 by
\bsax are explained in the framework of the IC model.  
In particular, these results appear to be related to
the radio spectral index structure of the
radio halos or relics present in these clusters.  
The essential requirement to detect nonthermal X-ray emission in the PDS
energy range is  
the presence of large regions of reaccelerated electrons,
with $\gamma$ $\sim$ 10$^4$, due to the
balance between radiative losses and reacceleration
gains in turbulence generated by recent merger events.

\section{Acknowledgments}

We thank the referee for valuable comments and suggestions.

\newpage

\figcaption[a3667.eps]{PDS data. The continuous line represents 
a thermal component at the average cluster gas temperature of 
7 keV (Markevitch \etal 1998). The errors bars are quoted at 1$\sigma$ level.}

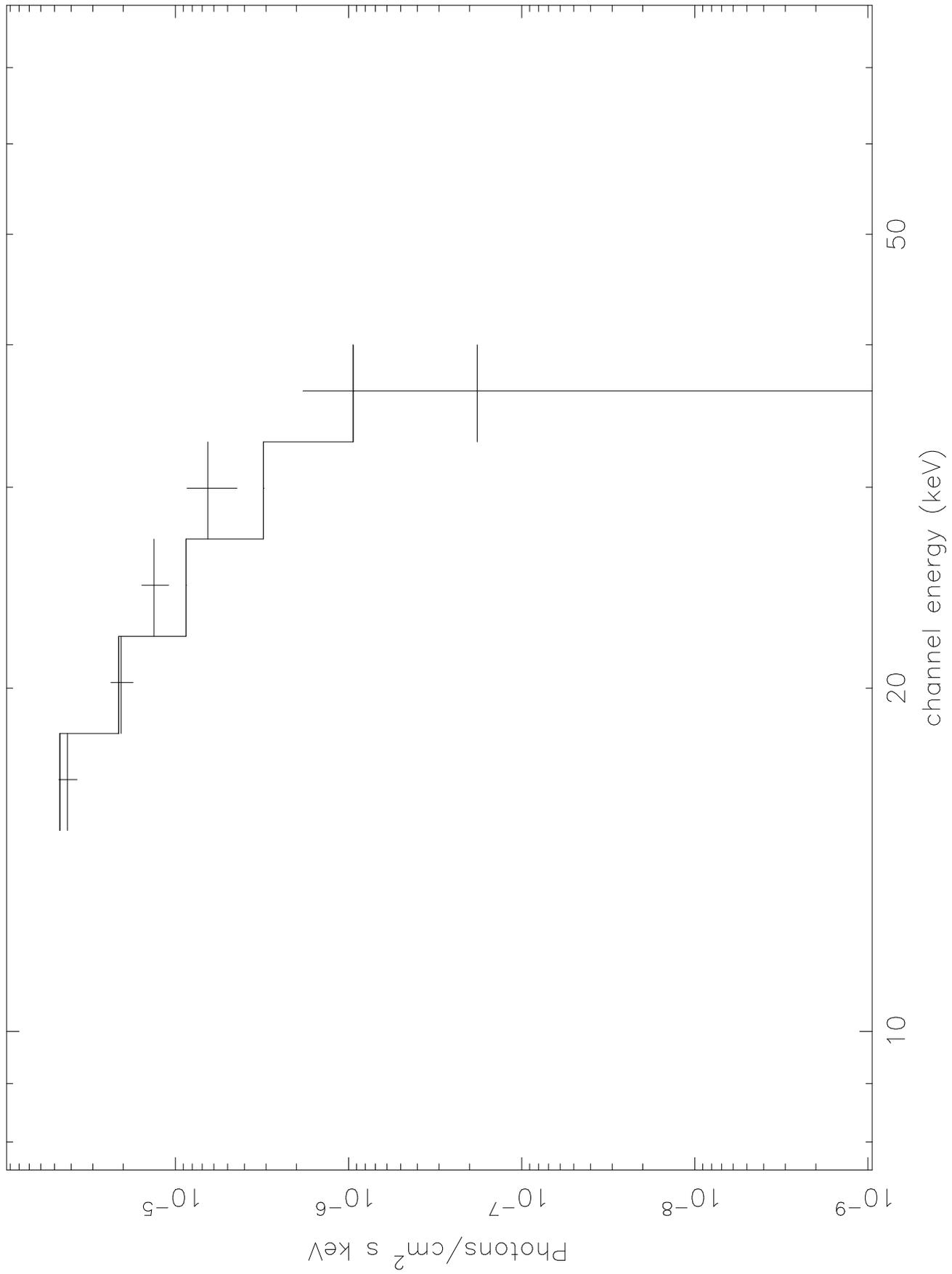
\end{document}